\documentclass[final,5p,times,twocolumn]{elsarticle}
\usepackage{graphicx}

\begin{document}

	\begin{frontmatter}
		\title{Hubbard bands, Mott transition and deconfinement in  strongly correlated systems
		}
		
		
		
		\author{V. Yu. Irkhin
			}
		
		
		
		\address{M. N. Mikheev Institute of Metal Physics, 620108 Ekaterinburg, Russia}
		\address{Ural Federal University, Ekaterinburg, 620002 Russia}
		\begin{abstract}
			{
								The problem of deconfinement phases in strongly correlated systems is discussed. In space-time dimension $d=3+1$, a competition of confinement and Coulomb phases occurs, but in $d=2+1$ the confining phase dominates owing to monopole proliferation, but gapless fermion excitations can change the situation. Combining the Kotliar-Ruckenstein representation and fractionalized spin-liquid deconfinement picture,  the Mott transition and  Hubbard subbands are treated,  general expressions in the case of an arbitrary bare band spectrum  being obtained. The transition into a metallic state is determined by condensation of a gapless boson mode. The spectrum picture in the insulating state is considerably influenced by the spinon spin-liquid spectrum and  hidden Fermi surface.
				
			}
		\end{abstract}
	\end{frontmatter}

\textbf{1. Introduction}
\newline

The  problem of Mott (metal-insulator) transition \cite{Mott} is very old, but is still of interest and importance. Usually this transition occurs in antiferromagnetic phase (Slater scenario, see \cite{Katsnelson,Igoshev}), but the situation changes for frustrated  systems: only the paramagnetic metallic and insulator  state are involved, a spin liquid being formed \cite{Vojta,Senthil2}. The transition into insulator state is related to correlation  Hubbard splitting (the Mott scenario). In the Mott state the  gap in the spectrum  is essentially the charge gap determined by boson excitation branch. Therefore the electrons become fractionalized: the spin degrees of freedoms are determined by neutral fermions (spinons), and charge ones by bosons. The corresponding slave-boson representation was first introduced by Anderson  \cite{Anderson,Wen}. 

In fact,  boson and fermion are coupled by a gauge field,  so that the problem of confinement occurs \cite{Wen}. The  transition into the metallic confinement state is described as a Bose condensation, the electron Green's function acquiring the finite residue.  On the other hand, in the insulator state the bosons have a gap, so that the spectrum is incoherent (the electron Green's function is a convolution of boson and fermion ones) and includes  Hubbard's bands. 

New theoretical developments provided a topological point of view for the Mott transition, since spin liquid possesses topological order (see reviews in \cite{Wen22,Scr2}). Phase transitions in  frustrated  systems (e.g., in a triangular lattice \cite{Wintel}) can be treated in terms of topological  excitations (instantons, monopoles, visons, vortices) which play a crucial role for confinement. 

In the present work we analyze the metal-insulator transition from the topological point of view in the situation of spin-charge separation. In Sect. 2 a general picture of confinement and deconfinement in a gauge field theory is discussed.
In Sect. 3 we adopt the deconfinement picture and apply the Kotliar-Ruckenstein slave-boson representation to calculate the spectrum of Hubbard subbands and investigate the Mott transition into the spin-liquid state, the relation of charge  gap in the boson spectrum and Hubbard splitting being established.
\newline

\textbf{2. Gauge field and vortices: Coulomb and Higgs phases in the Bose Mott insulator}
\newline

In a  general case of a correlated electron system we have a material field  interacting with  an effective gauge field,
its singularities  determining the monopole physics. Fluctuations of the gauge field can be treated as fluctuations of spin chiralty, which are important for the transport properties of Mott systems.

A useful analogy is given by the charged Bose liquid in a magnetic field where one has to take into account the gauge invariance. Here, the magnetic field penetrates into the  sample as vortex filaments which carry unit flux quanta and can originate and terminate at instantons and anti-instantons \cite{Fradkin,Nagaosa1}.
Then the quantization of the  magnetic  flux and  the Meissner effect  occur, similar to situation in a superconductor. 

In the pure gauge model, the   magnetic monopoles (instantons) -- the point singularities of the gauge field -- occur, which are  non-local excitations of the system, interacting according to the Coulomb law.
In the case of compact field, a confinement situation can occur owing to the monopoles.

With adding a material field (in the simplest model, a scalar Bose field) interacting with the gauge field, the situation is more complicated: the Higgs effect (occurrence of the gauge boson mass) can result in formation of a new ``Coulomb'' phase which is essentially a deconfined phase. 


A formal theory starts from the Anderson-Higgs action for  the gauge field  interacting 
with a material field \cite{Fradkin,Kogut,Nagaosa1,Wen},
\begin{eqnarray}
	{\mathcal S} = - \beta \sum_{\rm link } (1-\cos [ \Delta_\mu \theta (i) -  q a_{\mu}(i) ] )
	\nonumber \\
	+ { 1 \over g} \sum_{\rm plaquette} \cos
	[ \Delta_\mu a_\nu (i) - \Delta_\nu a_{\mu}(i) ].
\end{eqnarray}
Here $i$ is the space-time lattice point, and  $\mu$, $\nu$ determine the direction in the lattice. the difference operator $\Delta_\mu$ reads
$\Delta_\mu f(i) = f( i + \mu) - f(i),$
the angular continuous variable $\theta(i)$ and the field $a_\mu(i)$ are compact and defined on a plane in the interval $[0, 2\pi]$,
$q$ is an integer charge,
$g$ is the coupling constant
of the gauge field, the constant $\beta$ determines coupling to the material field 
and is related to the Higgs length $R$, $\beta = 1/R^2$.
The second term in the action (1) can be represented in terms of potential energy (fluxes of electric charge) and kinetic energy (effective magnetic field) \cite{Kogut}. 
A  picture of  confinement can be qualitatively explained as competition of these  energies  \cite{Kogut,Nagaosa1,Scr2}. 
In the case of a strong coupling $g$, only the electric term is important, the energy of the closed loop of the flux being proportional to its length. The magnetic term in this case makes possible for loops to fluctuate, but always leaves them closed. For small $g$, the gauge field is deconfining.
The conservation of the flux in the topological phase with a non-compact field means the conservation of the charge, and consequently the presence of bosons.


The  pure gauge model in $d =2+1$ is always confining at arbitrarily small coupling constant \cite{Polyakov} owing to occurrence of instantons 
which provide tunneling events.
In the presence of a material field, the situation can change due to the Higgs phenomenon. 
The phase diagram for $d = 3+1$  case contains the Higgs-confinement 
phase and Coulomb (free charge, small $g$) phase, see Fig. 1  of Ref. \cite{Nagaosa1}.  
A  crossover between the Higgs and confinement states is also possible.
In the Coulomb phase, the gauge field is deconfining and massless, and the Bose field remains disordered.

On the other hand, the phase diagram for $d =2+1$ was debatable.
For small $g$, where gauge fluctuations are weak,  the system reduces to the XY-model weakly coupled to a U(1) gauge field, which yields
an ordered Higgs phase at zero temperature. 
In the Higgs phase, the gauge field acquires  a gap by the
Anderson-Higgs mechanism. It is
also gapped in the confinement phase due to the screening
of magnetic charges \cite{Wen}. 
According to papers \cite{Einhorn,Einhorn1}, which considered 
the free energy of the finite vortex segments,
the XY transition remains at least for weak $g$, so that a finite region of the
Coulomb phase occurs.
However, including the effect of instantons for general $\beta$ and $g$ values leads to a different result \cite{Nagaosa1}:
the Higgs-confinement phase is better described as the confinement phase rather than the Coulomb one, with the exception of the $g=0$ line where the XY-phase transition takes place.
Thus for $d=2+1$ we have only the confinement phase where the gauge field is massive due to instantons.

In the strong coupling  (large $g$) limit the gauge field does not have its own dynamics and provides the constraint of integer boson occupation at each lattice site,
resulting in an insulator state. 
Therefore the confinement phase may be understood
as a Bose Mott insulator.
This Mott phase turns out to extend to include the entire phase diagram. 

The insulator-to-metal transition is essentially a
condensation of charged bosons coupled to a gauge
field \cite{Senthil2,Vojta}.

The appearance and picture of the topological excitations is explained in Ref. \cite{Einhorn}.  The pure XY model without gauge potential has for $d=2+1$ topological singularities given by closed vortex loops which are represented by a conserved topological current. 
The topological singularities of the abelain Higgs
model are a combination af  closed vortex loops and open strings
terminating on monopoles, and there are no free monopoles.  
In the pure gauge theory for $d=3+1$, the singularities 
are monopole loops, i.e., ``world lines'' of monopole-antimonopole pairs. 
In the XY model in $d=3+1$, the currents are conserved,
and represent closed orientable surfaces (spheres and spheres with 
handles). 
In addition to these closed surfaces, 
this model  has excitations which
are slices through these surfaces (or windows on
the surface) which are bounded by the monopole
loop. In $d=2+1$ current lines terminate
on monopoles, and in $d=3+1$ a surface
terminates on monopole loops. 
Thus  the topological excitations of
this model  are closed ``vortex''
surfaces of dimension $d- 2$, and open vortex surfaces of dimension $d- 2$ bounded by monopole surfaces of dimension $d- 3$ \cite{Einhorn1}.



In the consideration of two-band model \cite{Sachdev}, the Fermi liquid (FL) state corresponds to the Higgs phase, and the U(1) fractionalized Fermi liquid (FL$^*$) state to the Coulomb phase, the monopoles playing no essential role.
Indeed, the information on the compactness of the U(1) gauge field is lost,
since it becomes effectively non-compact,
so that monopole excitations are suppressed. 
In the U(1) $d=3+1$ FL$^*$ phase, which is an analogue of the Coulomb phase
of the compact gauge theory, the monopole gap is finite. 
On the other hand, in the FL phase, the
monopoles do not exist being confined to each other 
due to the condensation of the boson field.
In such a model, transition from small Fermi surface in FL$^*$ phase to large Fermi surface in FL phase occurs.
A similar description of the interaction- and doping-induced Mott transition can be obtained \cite{Senthil2,Senthil1}.

The compactness situation   in $d=2+1$ can change in a gapless spin liquids with a large number of gapless fermionic matter 
fields 
\cite{Senthil2}. 
According to \cite{Hermele},
in the problem of a Fermi surface of spinons
coupled to a compact U(1) gauge field in two
dimensions, one can expect 
that monopoles are irrelevant for large number of Dirac points $N$. Thus the monopole-free theory is  sensitive to Fermi surface instabilities. 
The evolution of the phase diagram in the SU(N) model depending on $N$ 
has been recently investigated  by using the Monte Carlo method \cite{Da}.
In  $d=3+1$, the
large-$N$ limit does not play a role: 
U(1) spin liquids can exist as stable phases in $d=3$ even
if the spinons are completely gapped \cite{Sachdev,Hermele}. 
Note that in $d=2+1$, the state of  Z$_2$ spin liquid can occur, which is unstable with respect to superconductivity \cite{Fradkin,Wen}.


According to \cite{Volovik}, the Fermi surface is a   singularity of the Green's function, which is characterized by a  topological invariant and  is protected by
topology. The situation is similar to the theory of superfluid liquid which contains peculiar   lines -- vortex filaments. 
The  vortex line in the frequency-quasimomentum space cannot be destroyed by small perturbations.  
This treatment can be supposed to be relevant also for strongly correlated systems,  in particular  for Mott insulators where  
the  Fermi surface is expected to be preserved even in the insulating phase.
The conservation of the  Fermi surface at a quantum phase transition is supported by existence of a spinon Fermi surface in the paramagnetic  Mott insulator phase \cite{Sachdev,Senthil2}. 
Thus in the gapped (e.g., Mott phase) the usual Fermi surface does not exist, but is transformed and becomes  ghost (hidden). Then the Luttinger theorem (the conservation of the volume enclosed by Fermi surface) is still valid \cite{Volovik}.
This idea was also applied to a half-metallic ferromagnet \cite{Scr3}.



An analogy with the above consideration of the gauge field with account of electric charge can be mentioned.   
In the case of non-compact gauge field the  picture corresponding to Refs. \cite{Volovik,Volovik1}  remains valid: e.g., for $d=3+1$ we have the  Coulomb phase. However, for the compact field (e.g., for $d=2+1$) we have tunneling events (monopole proliferation)  resulting in confinement,
vortex segments being broken into pieces, see Fig. 2 of Ref. \cite{Nagaosa1}.
In such a way, we can expect instabilities of the Fermi surface.
\newline

\textbf{3.  Mott transition and Hubbard bands in the Kotliar-Ruckenstein representation}
\newline

The Mott transition can be interpreted as a
softening of an auxiliary Bose excitation. 
To describe this transition, a number of works \cite{Senthil2,Kim,Lee,Hermele1} applied the rotor representation which is simple, but not quite convenient, since it does not include explicitly the spectrum of both Hubbard bands.
An alternative description of the  transition and Hubbard bands  can be obtained by using the Kotliar-Ruckenstein slave-boson representation \cite{Kotliar86}.
This problem was considered in \cite{lavagna,raimondi}
for usual paramagnetic state by using a Gutzwiller-type approach. Here we propose a more advanced treatment taking into account spin-liquid formation.

The representation \cite{Kotliar86} uses the Bose operators $
e_{i},\,p_{i\sigma },\,d_{i}$ and Fermi  operators $f_{i\sigma }$: 
\begin{equation}
	c_{i\sigma }^{\dag }\rightarrow f_{i\sigma }^{\dag }z_{i\sigma }^{\dag
	},~z_{i\sigma }^{\dag }=g_{2i\sigma }(p_{i\sigma }^{\dag
	}e_{i}^{{}}+d_{i}^{\dag }p_{i-\sigma }^{{}})g_{1i\sigma },
\end{equation}%
with the constraints
\begin{equation}
	\sum_{\sigma }p_{i\sigma }^{\dagger }p_{i\sigma }+e_{i}^{\dagger
	}e_{i}+d_{i}^{\dagger }d_{i}=1,~\,f_{i\sigma }^{\dagger }f_{i\sigma
	}=p_{i\sigma }^{\dagger }p_{i\sigma }+d_{i}^{\dagger }d_{i}  \label{con},
\end{equation}%
the factors  $g_{1,2i\sigma}$ being somewhat arbitrary, but in the mean-field approximation for a non-magnetic state we can put $g_{1,2\sigma}^{2}=2$ \cite{raimondi}. Also in this approximation we can put $p_{i\sigma }^{2}=1/2$. 
The corresponding Lagrangian of the Hubbard-Heisenberg model reads
\begin{eqnarray}
	\mathcal{L} &=&-\sum_{ij\sigma }t_{ij}z_{i\sigma }^{\dag }z_{j\sigma
	}f_{i\sigma }^{\dag }f_{j\sigma }+\sum_{i\sigma }f_{i\sigma }^{\dag
	}(\partial _{\tau }-\mu +\lambda _{2\sigma })f_{i\sigma }  \nonumber \\
	&+&\sum_{i\sigma }p_{i\sigma }^{\dag }(\partial _{\tau }+\lambda
	_{1}-\lambda _{2\sigma })p_{j\sigma }+\sum_{i}e_{i}^{\dag }(\partial _{\tau
	}+\lambda _{1})e_{i}  \nonumber \\
	&+&\sum_{i}d_{i}^{\dag }(\partial _{\tau }+\lambda _{1}-\sum_{\sigma
	}\lambda _{2\sigma }+U)d_{i}+\sum_{ij }J_{ij}{\bf S}_i{\bf S}_j.  \label{lag}
\end{eqnarray}%
Although the Heisenberg interaction is obtained in the Hubbard model as an effective superexchange interaction (in the second order of perturbation theory), for convenience it is explicitly included in (\ref{lag}).
In the mean-field approximation the Lagrange factors $\lambda_{1,2} $
corresponding to (\ref{con}) are site-independent. In the insulator phase 
we have \cite{lavagna} $\lambda _{1 }=\lambda _{2 \sigma} = U(1 \pm\zeta)/2$ being the chemical potential for an infinitesimally small electron (hole) doping (added or removed particle),  $\zeta =(1-U_{c}/U)^{1/2}$. Here 
$$U_{c}=4p^2 g_{1}^{2}g_{2}^{2}\varepsilon=8 \varepsilon $$ 
is the critical value for the Mott transition
in the Brinkman-Rice approximation,  $\varepsilon =2\left\vert \int_{-\infty }^{\mu }d\omega \omega \rho
(\omega )\right\vert $
the average non-interacting energy, $\rho (\omega )$ the bare density of
electron states.

The calculation of the  Green's function for the boson combination  $b_{i}^{\dag }=e_{i}^{\dag }+d_{i}$ yields (cf. Ref.\cite{raimondi})
\begin{eqnarray}
	D(\mathbf{q},\omega ) &=&\langle \langle b_{\mathbf{q}}|b_{\mathbf{q}}^{\dag
	}\rangle \rangle _{\omega }=\sum_{a =1,2}\frac{Z_{\alpha \mathbf{q}}}{
		\omega -\omega _{\alpha \mathbf{q}}},~ \\
	Z_{a \mathbf{q}} &=&(-1)^{a}U/\sqrt{U^{2}\zeta ^{2}+U (U_{c}- 4\Sigma ({\mathbf{q}}))}
\end{eqnarray}
where the spectrum of boson (holon-doublon) subsystem is given by
\begin{eqnarray}
	\omega _{a \mathbf{q}} &=&\frac{1}{2}[\pm U\zeta - (-1)^a \sqrt{U^{2}\zeta		^{2}+U (U_{c}- 4 \Sigma ({\mathbf{q}}))}] \label{la}
\end{eqnarray}%
Here we have taken into account the boson self-energy 
\begin{equation}
	\Sigma (\mathbf{q})=-p^2 g_{1}^{2}g_{2}^{2}\sum_{\mathbf{k}\sigma }t_{\mathbf{k-q}}n_{\mathbf{k}\sigma },\, n_{\mathbf{k}\sigma}=\langle f_{\mathbf{k\sigma }}^{\dag }f_{\mathbf{k\sigma 
	}}^{ }\rangle, \label{SF}
\end{equation}%
which is obtained by a simple decoupling of the first term in (\ref{lag}) and is essentially the Harris-Lange correction \cite{Harris}. 

The dispersion of bosons is influenced by details of fermion spectrum which is determined by the  $f$-system state. 
Spin degrees of freedom can be treated separately with the Heisenberg Hamiltonian in the $f$-pseudofermion representation. Under some conditions, one can expect formation of a spin-liquid state where excitation are essentially spinons -- neutral fermions.

The mean-field picture of spinon spectrum $E_{\bf k}$ can be stabilized in the case of a non-compact gauge field or by gapless Fermi excitations. In the insulator state this spectrum is not influenced by bosons, various spin-liquid phases being obtained \cite{Wen}. For a square lattice, in the uniform RVB (uRVB) phase $E_{\bf k} \sim J (\cos k_x+\cos k_y)$. In the $\pi$-flux ($\pi$Fl) phase (which contains Dirac points) $E_{\bf k} \sim \pm J \sqrt{\cos^2 k_x + \cos^2 k_y}$; this spectrum is obtained using the SU(2) invariance of the Heisenberg interaction. Also, gapped Z$_2$ phases can occur if the next-nearest-neighbor interaction is present.

In the absence of considerable $\mathbf{k}$-dependence of $n_{\mathbf{k\sigma }}$ (a localized spin phase without fermion hopping), $\Sigma$ tends to zero. However, for a spin liquid we have a sharp Fermi surface. 
Although, generally speaking, the spinon spectrum form differs from bare electron one,
for $q=0$ we still have $\Sigma(0)=U_c/4$  since the spinon band is  half-filled and the chemical potential (the position of the Fermi energy) is fixed.

In the nearest-neighbor approximation, after passing in (\ref{SF}) to the coordinate representation one can  see that the spectrum of spinons and correction to holon spectrum differ, roughly speaking, only in the replacement of $J$ by $t$ ($\Sigma({\bf q}) \propto E(\bf{q})$, cf. Ref. \cite{Wen} for the $t-J$ model).
In particular, we have
\begin{eqnarray}
	\Sigma (\mathbf{q})=U_c(\cos q_x+\cos q_y)/8, \nonumber \\
	\Sigma (\mathbf{q})=\pm U_c\sqrt{\cos^2 q_x + \cos^2 q_y}/(4\sqrt{2})
\end{eqnarray}%
for the uRVB and $\pi$Fl phases, respectively.
In the large-$U$ limit we have $$\omega_{a\mathbf{q}}= {\rm const} - (-1)^a \Sigma({\bf q})/\zeta.$$

We stress that our interpretation of spectrum is different from that in Ref. \cite{raimondi} where the limit of vanishing renormalized electron bandwidth (i.e., in the Mott phase where the averages $e,\,d\rightarrow 0$) is treated in a Gutzwiller-type approach.
It is important that a characteristic scale of spinon energies is small in comparison with that of electron ones, so that the spinon Fermi surface is strongly temperature dependent; this situation is somewhat similar to the case of magnetic order.
Note also that  for magnetically ordered phase such approach does not work since the above approximation for factors $g$  is not valid \cite{Irk}.






At the critical point  we should have 
\begin{equation}
	{\rm Im}D(\mathbf{q},\omega )  \propto \frac{\theta\left(\omega^2 - c^2q^2\right)}{\left(\omega^2 - c^2 q^2\right)^{(2-\eta)/{2}}} 
\end{equation} 
where $\eta$ is the anomalous critical exponent of the boson field at the $3D$ XY fixed point \cite{Fisher}.
Since $\eta $ is small,  the bosonic
sector behaves roughly as in the absence of gauge
field while the spinon spectrum is considerably modified by this \cite{Senthil2}. 

The observable electron Green's function is obtained as a convolution of the boson and spinon Green's functions.
Since as a rule $J \ll  |t|$, an account of this spinon smearing (unlike of the boson dispersion, which is also related to spinon energies) does not strongly influence the shape of density of state. Then we can put ${\rm Im} \langle \langle f_{\mathbf{k\sigma }}^{ }|f_{\mathbf{k\sigma }}^{\dag }\rangle \rangle _{E} \sim \delta(E-\lambda_2)$ to obtain the Hubbard bands with the energies $\lambda_2 - \omega_{1,2 \mathbf{q}}$ for vanishing electron (hole) doping with energies near 0 and $U$, respectively, $\lambda_2 $ being the corresponding chemical potential \cite{raimondi}.
This spectrum corresponds to upper and lower Hubbard subbands with the width of order of bare bandwidth, the gap vanishing  at the transition point $U\rightarrow U_c$. 

In fact, spinons are not fully free. 
At some points in the Brillouin zone the interaction with the gauge field owing to constraint $\sum_{\sigma }f_{i\sigma }^{\dagger }f_{i\sigma }^{ }=1$ (which is approximately satisfied owing to (\ref{con})) can play an important role. In the leading order in $1/N$ (e.g., for large number of Dirac points $N$), one has for the spinon field $ \psi ({\bf r},t)$  \cite{Rantner,Wen} 
\begin{equation}
	\langle \psi ({\bf r},t)\psi ^{\dagger }(0)\exp \biggl(i\int_{0}^{x}\sum_{\mu=0,1,2} a_\mu dx_\mu \biggr)\rangle
	\propto \frac {1}{({\bf r}^{2}+t^2)^{1-\alpha /2}}
\end{equation}%
with anomalous dimension $\alpha =32/3\pi ^{2}N$ ($N=2$) at the $(\pi/2,\pi/2)$ point. Here  $x=({\bf r},t)$, $\int_{0}^{x}dx$ is
the integration along the straight return path and $\langle...	\rangle$ includes integrating out the gauge fluctuations.
Then the observable electron Green's function, being again obtained as a convolution,  demonstrates a singular non-quasiparticle  behavior \cite{Rantner}:
\begin{equation}
	G_{e}({\bf r},t) \propto \langle b^{\dagger }({\bf r},t)b(0)\rangle _{0} 
	\frac {1}{({\bf r}^{2}+t^2)^{1-\alpha /2}}
\end{equation}%
where  the subscript 0 means neglecting gauge fluctuations.
\newline

\textbf{3. Discussion}
\newline

Field-theoretical approaches including a gauge field enable one to describe efficiently  formation of the correlated Hubbard bands.
In the deconfinement conditions, the structure of spectrum is strongly influenced by the spinon excitations and should demonstrate a strong temperature dependence. The expressions for the Green's functions obtained can be  used to calculate the optical conductivity, cf. Ref. \cite{raimondi}.

As discussed in Sect. 2, the boson picture can provide integer site occupation,  i.e., Hubbard's projection.
As well as the consideration of  the chiral spin liquid, where the physical quasiparticles are spinons dressed by a $ \pi $ vortex \cite{Wen22,Scr2}, the mutual Chern-Simons  theory \cite{Weng} enables one to realize the no-double-occupancy constraint.
Here, the quantization of the flux of the gauge fields with unit value of $\pi$ results in integer values of spinon and holon occupation numbers.



Although most theoretical investigations are performed in $d=2+1$, spin-liquid states can occur in some three-dimensional systems, e.g., pyrochlores \cite{pyrochlores,Wan}. Even if an instability with respect to magnetic ordering or superconductivity occurs in the ground state,  a spin-liquid-like state can occur in an intermediate temperature regime \cite{Sachdev}, especially in frustrated systems.

We see that the picture of Mott transition into a non-magnetic ground state is connected with topological features: the deconfined spin-liquid state involved includes fractionalization and long-range many-particle quantum entanglement \cite{Scr2}. Generally, description of the correlated paramagnetic phase, which may have a complicated internal structure, is an important problem. 
Further developments are  connected with the string condensate theory
\cite{Levin} where the lines of the effective electric flux, discussed in Sect. 2, determine strings. The details of the string picture are determined by the coupling constant $g$. 
At large $g$, potential energy dominates, so that the closed loops of the electric flux become constricted.
On the contrary, at small $g$ the kinetic energy, whose role is played by the magnetic field, prevails, so that many loops are present.
The corresponding deconfinement phase is the quantum spin liquid --  the condensate of large strings, and fermions appear as ends of
open strings in any dimension.

Recently, a number of investigations of the Mott insulator  problem including spin-liquid states have been performed, various  scenarios of deconfinement and  Fermi surface instabilities 
being proposed \cite{Chowdhury,Song,You}.
In the model \cite{Song}, two scenarios of the transition from the Mott insulator (chiral spin liquid) can arise. If holons condense, a chiral metal with enlarged unit cell and finite Hall conductivity is obtained. In a second  scenario, the internal magnetic flux adjusts with doping and holons form a bosonic integer quantum Hall state which is identical to a $d + id$ superconductor.
The transition of holons from Mott insulator or the  Hall state
to superfluid is described by Bose condensation in $d=2 + 1$, so that the electron Fermi surface arises.

The paper \cite{You} treated the SO(4) bosonic topological transition. The corresponding   model is defined on the double layer honeycomb lattice and includes spin-orbit coupling.  
There occurs a featureless Mott phase  and two bosonic symmetry-protected topological  phases which are separated by quantum phase transitions. A fermionic transition corresponds to the Dirac semimetal  and
two bosonic topological transitions. This example demonstrates disappearance of the Fermi surface at the topological transition.

The author is grateful to Yu. N. Skryabin for numerous fruitful discussions.
The research funding from the Ministry of Science and Higher Education of the Russian Federation (the state assignment, theme ``Quantum'' No. 122021000038-7, and Ural Federal University Program of Development within the Priority-2030 Program) is  acknowledged.



\end{document}